\documentclass[prl,twocolumn,showpacs,preprintnumbers,amsmath,amssymb,floatfix,superscriptaddress]{revtex4}

\usepackage[]{graphicx}         	
\usepackage{bm}             		
\usepackage{tabularx}

\begin{document}

\title{Current-driven vortex oscillations in metallic nanocontacts}

\author{Q. Mistral}
\affiliation{Institut d'Electronique Fondamentale, UMR CNRS 8622, Universit{\'e} Paris-Sud, 91405 Orsay, France}
\author{M. van Kampen}
\affiliation{IMEC, Kapeldreef 75, B-3001 Leuven, Belgium}
\author{G. Hrkac}
\affiliation{Department of Engineering Materials, University of Sheffield, Sheffield S1 4DU, United Kingdom}
\author{Joo-Von Kim}
\email{joo-von.kim@ief.u-psud.fr}
\author{T. Devolder}
\author{P. Crozat}
\author{C. Chappert}
\affiliation{Institut d'Electronique Fondamentale, UMR CNRS 8622, Universit{\'e} Paris-Sud, 91405 Orsay, France}
\author{L. Lagae}
\affiliation{IMEC, Kapeldreef 75, B-3001 Leuven, Belgium}
\author{T. Schrefl}
\affiliation{Department of Engineering Materials, University of Sheffield, Sheffield S1 4DU, United Kingdom}

\begin{abstract}
We present experimental evidence of sub-GHz spin-transfer oscillations in metallic nano-contacts that are due to the translational motion of a magnetic vortex. The vortex is shown to execute large-amplitude orbital motion outside the contact region. Good agreement with analytical theory and micromagnetics simulations is found.
\end{abstract}


\maketitle

Lateral confinement in magnetic nanostructures leads to the appearance of complex magnetic states. One pertinent example concerns the magnetic vortex, which is well-studied and has gained renewed interest due to improved lithography and fabrication techniques. The dynamics of such topological objects is, in part, governed by the confining potential. For instance, magnetic vortices possess a translational mode by which the vortex core executes a spiralling motion about its equilibrium position~\cite{Guslienko:JAP:2002,Ivanov:JAP:2004}. Recent experimental studies have demonstrated resonance phenomena with such cores~\cite{Novosad:PRB:2002b,Park:PRB:2003,Buchanan:NatPhys:2005,Novosad:PRB:2005,Guslienko:PRL:2006}, and proposals have been made for magnetic resonators based on these objects.

More recently, it has been shown experimentally that such vortices can be brought into a self-oscillatory state~\cite{Pribiag:NP:2007}. This has been achieved with the spin-transfer effect in which a spin-polarized current transfers spin-angular momentum to the magnetic structure, giving rise to an additional torque exerted on the magnetization~\cite{Slonczewski:JMMM:1996,Berger:PRB:1996}. Under certain conditions the spin-transfer compensates the natural dissipation processes, leading to self-sustained oscillations. In contrast to a resonator in which oscillations are driven by a periodic external force, the self-oscillator here is driven by a constant electric current only. While current-driven oscillations related to spin-waves in magnetic multilayers are well-studied~\cite{Tsoi:Nature:2000,Kiselev:Nature:2003,Rippard:PRL:2004}, Ref.~\onlinecite{Pribiag:NP:2007} provides the first conclusive evidence of this phenomenon for a magnetic vortex.

In the context of spin-transfer, confinement can also arise from the distribution of applied current itself. One pertinent example is the metallic point-contact in which the current is applied through a small metallic cross-section in contact with a continuous magnetic film~\cite{Rippard:PRL:2004}. In the absence of any applied currents, a vortex can exist in a continuous film but, aside from sample edges and point defects, there is no magnetic potential in which vortex oscillations can take place. In the presence of an applied current, however, the Oersted fields produce an attractive potential centered at the contact center. Earlier experimental studies have provided hints at the existence of vortex modes, but no explanations for the observed low-frequency excitations were given~\cite{Pufall:PRB:2007}. In this Letter, we show that the cylindrical nature of the Oersted fields, generated by the applied current, is favorable to the creation of a magnetic vortex and provides a potential in which the vortex oscillates. Moreover, we demonstrate that the sub-GHz frequencies of these vortex self-oscillations result from a vortex orbit that is actually \emph{outside} the contact region.

The magnetic structure we have studied in experiment is a metallic point contact deposited on a metallic spin-valve stack. The multilayer is composed of Ta(3.5 nm) / Cu(40) / Ta(3.5) / Ni$_{80}$Fe$_{20}$(3) / IrMn(7) / Co$_{90}$Fe$_{10}$(3.5) / Cu(3) / Ni$_{80}$Fe$_{20}$(4) / Pt(3). The NiFe (4 nm) layer is the magnetic free layer and the CoFe layer, which is exchange biased by the IrMn antiferromagnet, serves as a reference layer for the giant magnetoresistance and spin-transfer. The layers are sputter-grown in an ultrahigh vacuum system with a base pressure of $3 \times 10^{-8}$~Torr. The metallic point contacts are made to the laterally extended ($>$ 5 $\mu$m) stack which is incorporated in the central conductor of a coplanar waveguide (CPW) structure. After patterning by conventional lift-off lithography, the stack is passivated by a 50 nm SiO$_2$ layer deposited using rf sputtering. The point contacts are defined in a PMMA layer by electron-beam lithography and etched into the SiO$_2$ by a short dip in a buffered HF solution. Next, the CPW structure is deposited and patterned by lift-off lithography. Finally, the devices are annealed at 250~$^{o}$C for 10 minutes to improve the exchange bias coupling between the IrMn and the Co$_{90}$Fe$_{10}$ layers.

\begin{figure}
\includegraphics[width=7cm]{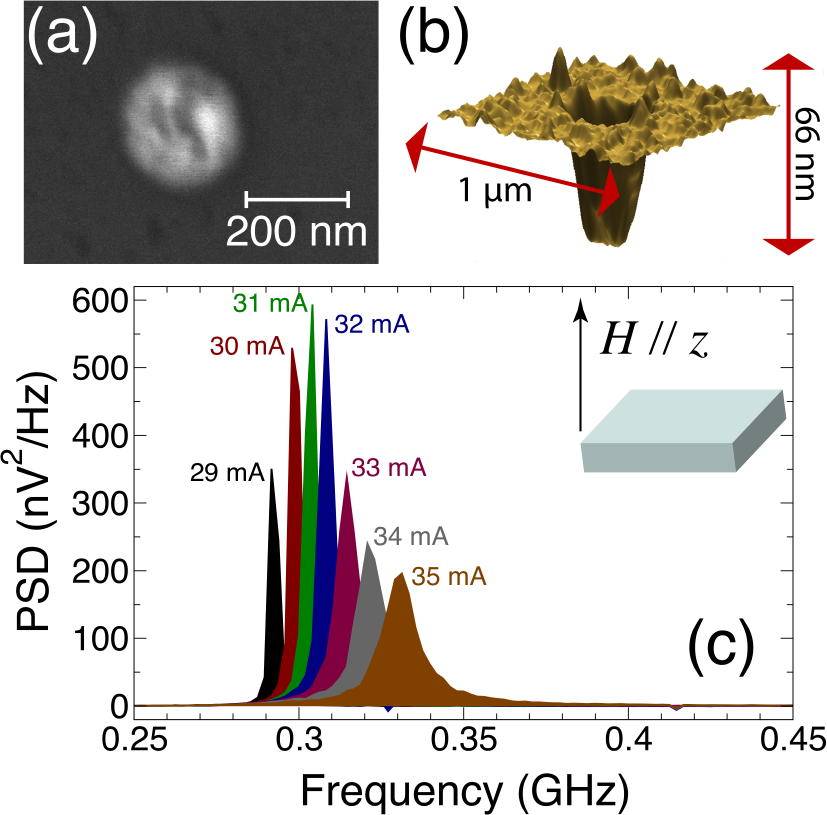}
\caption{\label{fig:spectra_contact}(Color online) (a) SEM and (b) AFM image of point contact. (c) Low frequency power spectra measured at $\mu_0 H = 0.21$ T for several applied currents. The field and film geometry is shown in the inset of (c).}
\end{figure}

The properties of the point contact have been characterized in detail by scanning electron microscopy (SEM), atomic force microscopy (AFM), ellipsometry and optical microscopy measurements. The target diameter of the point contacts are nominally 100 nm, however, due to the isotropic wet etch the diameter of the contacts is increased with respect to the 100 nm holes in the PMMA layer. Indeed, our measurements show that there is a significant variation in the diameter as a function of depth. In Fig.~\ref{fig:spectra_contact}a and b, we present an SEM and AFM measurement, respectively, of a typical point contact. While a diameter of 232 nm is extracted from SEM, AFM measurements show that the diameter can vary from 130 to 270 nm over a thickness of 40 nm. Based on the depth profile and our multilayer configuration, we estimate the contact diameter at the free layer to be approximately 200 nm.

For the magnetic properties, a uniaxial anisotropy field of $\mu_0 H_{k} =$ 1.4 mT for the free layer is determined from hard-axis hysteresis loop measurements. From the easy axis loops, we determine the biased layer switching to be around $-60$ mT. A small bias of 6 mT in the free layer switching is also observed and is attributed to N\'eel coupling between the magnetic layers. The saturation magnetization of the free layer is found to be $\mu_0 M_s = 1.1$ T from alternating gradient field magnetometer measurements. For the electrical measurements, the applied current perpendicular through the spin-valve stack is generated by applying a DC voltage to a $200~\Omega$ resistor in series with our sample. A 20~m$\Omega$ magnetoresistance is a typical figure for our spin-valve stacks. Two bias tees allow DC and RF routing. One of the RF routes is amplified (+43~dB over the 100~MHz - 26~GHz band) and connected to a spectrum analyzer. The other RF route is terminated by a 50~Ohm load while the DC route is shorted. The final power spectra data is obtained after subtracting a reference curve taken with a zero current bias. The rf measurements are made for magnetic fields applied perpendicular to the film plane, which improves the magnetoresistance signal because the projection of the free layer oscillations onto the fixed layer is larger. Positive currents describe electrons flowing from the fixed to the free layer.

An example of low-frequency power spectra is presented in Fig.~\ref{fig:spectra_contact}c. 
%
The peaks shown represent the fundamental mode for several applied currents under a perpendicular field of $\mu_0 H = 210$ mT. Harmonics to fourth order are also observed, but these are several orders of magnitude smaller than the fundamental, which indicates magnetization precession with small ellipticity. The oscillations have quality factors up to 250, which corresponds to the narrowest linewidth of 1.2 MHz observed at a frequency of 299 MHz. These low precession frequencies are suggestive of vortex motion rather than uniform precession~\cite{Novosad:PRB:2005}. Indeed, the Smit-Beljers equation~\cite{Smit:PRR:1955}  for uniform magnetization precession predicts a precession frequency of 1.08 GHz with our experimental parameters, which is clearly four to five times higher than what we observe. In contrast to earlier experiments in which nominally smaller diameter contacts were studied~\cite{Rippard:PRL:2004}, the Oersted fields in our samples are significantly larger (at equivalent current densities) at the contact edges and therefore favor vortex formation. For example, a 30 mA current generates an Oersted field of 60 mT at the contact edge, which is a significant fraction of the perpendicular applied fields we considered.

\begin{figure}
\includegraphics[width=7cm]{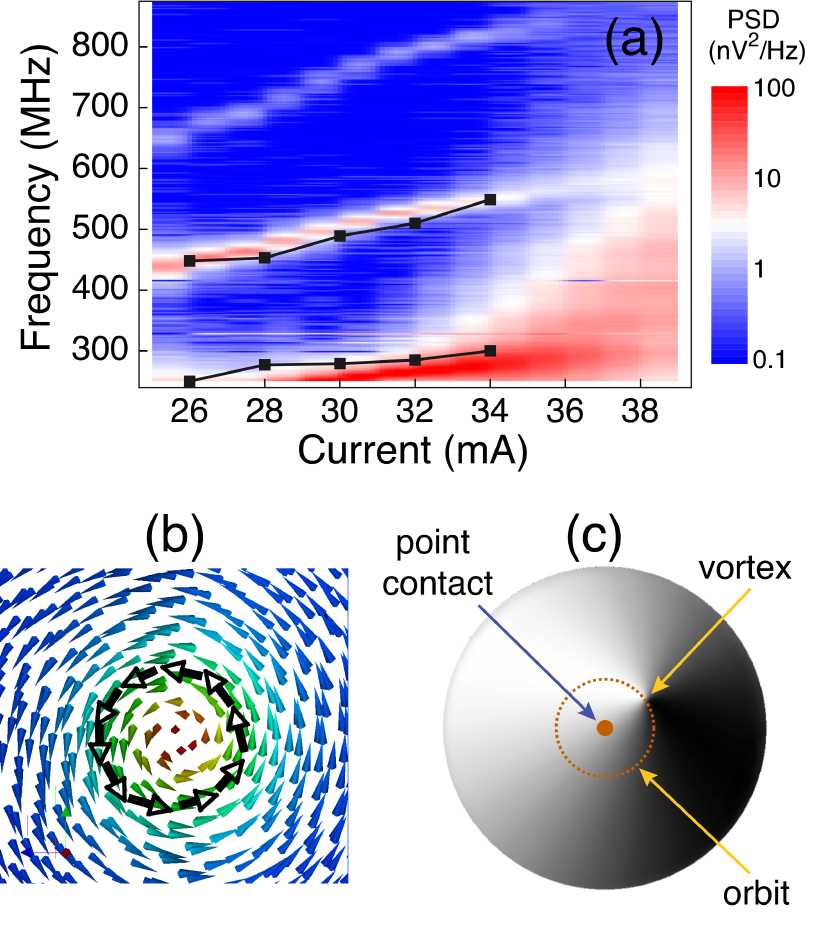}
\caption{\label{fig:map_simulation}(Color online) (a) Color map of experimental power spectral density (PSD) as a function of current for $\mu_0 H = $ 350 mT. Solid squares represent results of micromagnetics simulations. (b) Top view of vortex magnetization profile obtained from simulation. (c) In-plane component of magnetization represented in a graylevel (black for $m_y = +1$, white for  $m_y = -1$) for the entire simulation area. Also shown are the contact and the vortex orbit.}
\end{figure}

Two other key experimental observations support the hypothesis of vortex oscillations. First, the measured oscillation frequencies increase as a function of applied current for all applied fields considered ($\mu_0 H \leq 350$~mT). An example of this behavior is presented in Fig.~\ref{fig:map_simulation} in which the full power spectra is shown in a density plot.
%
This frequency blueshift is a feature of magnetization precession out of the film plane, but is not expected for the fields used in experiment. Indeed, the spin-torque oscillator theory of Ref.~\onlinecite{Slavin:ITM:2005}, for uniform magnetization precession, predicts a frequency \emph{redshift} for these applied perpendicular fields. Second, the precession mode is sensitive to small in-plane applied fields. The oscillations are observed to be stable for applied in-plane fields up to 15-20 mT in magnitude, and for larger fields the oscillations are completely suppressed. Furthermore, the narrowest peaks are obtained for in-plane fields that compensate the 6 mT bias field due to the N\'eel coupling, which is again consistent with the vortex picture.

To understand the exact nature of the low frequency oscillations, we performed full micromagnetic simulations of a trilayer CoFe(3.5 nm)/Cu(3)/NiFe(4) stack. The saturation magnetization of the free layer is taken to be the same as the experimental value, $\mu_0 M_s = 1.1$ T, and a GMR ratio of 1\% is used. We account for the inhomogeneous current distribution flowing through the free layer by computing the local current density from the local angle between the free and fixed layer magnetizations. The Oersted field is computed with the Biot-Savart law from this current distribution and an asymmetric Slonczewski term for the spin-transfer is used~\cite{Xiao:PRB:2004}. The simulation area is a circular film 1000 nm in diameter, with the 200 nm diameter point contact at the center. The system is discretized with a finite element method with a linear basis function and a discretization size of 4 nm, which is below the exchange length of 4.5 nm. An exchange bias field of 162 mT is applied to the reference CoFe layer to simulate pinning in the experimental stacks. 

The simulations are performed as follows. First, we calculate the remanent state with an external field applied perpendicular to the film plane and in the absence of currents. From an initial uniform state in the film plane, we allow the magnetization to relax toward the applied field direction through time-integration of the equations of motion. During this process, a vortex first forms under the point contact (which is situated in the center of the simulation grid) and is followed by the creation and annihilation of multiple vortices. After this transient phase, a single vortex state emerges as the system relaxes to its remanent state. Next, a current is applied through the  point contact with currents ranging from 26 mA to 36 mA, with the remanent state serving as the initial state for this step of the calculation. We observe that the additional spin torque drives the vortex out of the contact area and towards a stable orbit around the contact. 

These simulations show that the oscillations observed in experiment are related to a large-amplitude translational motion of a magnetic vortex. In contrast to the vortex dynamics observed in nanopillars in which the vortex core precesses within the confining part of the Oersted field~\cite{Pribiag:NP:2007}, the dynamics here correspond to an orbital motion \emph{outside} the contact region. This behavior can be likened to planetary orbital motion under the influence of a gravitational field; the spin-transfer torque leads to a centripetal motion of the vortex core, which is counterbalanced by the attractive potential provided by the Oersted field. Good quantitative agreement between the simulation and experimental frequencies is achieved, as shown in Fig.~\ref{fig:map_simulation}. In addition, the stability of the vortex has also been investigated using simulations. For small in-plane fields  (less than 10 mT for this system), a small blueshift in the frequency is observed. This leads to an asymmetry in the field profile relative to the vortex structure and results in an elliptical motion of the vortex orbit. We have also verified in simulation that larger in-plane fields lead to a breakdown of the vortex structure.

A qualitative understanding of how the oscillation frequencies change with applied field and current can be achieved in terms of a rigid vortex model. Let $\Theta(\vec{r})$ and $\Phi(\vec{r})$ denote the magnetization orientation in spherical coordinates, with the spatial variations in $\vec{r}$ given in polar coordinates $(\rho, \varphi)$ to exploit the symmetry of the circular point contact. We take the magnetization to be uniform across the film thickness. The vortex profile is approximated by a polar angle that only depends on the radial spatial coordinate, $\Theta = \Theta(\rho)$, and an azimuthal angle that depends only on the azimuthal spatial coordinate $\Phi = \varphi \pm \pi/2$, where the sign in the last expression denotes the vortex polarization. 

Based on a given magnetization profile, we can derive an equation of motion for the vortex core by following Thiele's method~\cite{Thiele:PRL:1973},
\begin{equation}
\vec{G} \times \frac{d \vec{X}_0}{dt} - \alpha \, \mathbf{D} \cdot \frac{d \vec{X}_0}{dt} + \sigma I \left(\vec{P}_\perp - \vec{P}_{||} \right)=  \frac{\gamma}{M_s} \frac{\partial W}{\partial \vec{X}_0},
\label{eq:thiele}
\end{equation}
where $\vec{X}_0 = (X_0,Y_0)$ represents the vortex core position in the film plane, $\gamma$ is the gyromagnetic ratio, $\alpha$ is the Gilbert damping constant, $\sigma$ is a measure of spin-transfer efficiency,
$\vec{G} = \int dV \sin{\Theta} (\nabla\Phi \times \nabla\Theta)$ 
is the gyrovector, 
$\mathbf{D} = \int dV \left( \nabla\Theta \bigotimes  \nabla\Theta + \sin^2{\Theta} \nabla\Phi \bigotimes  \nabla\Phi \right)$ 
is the damping dyadic, and $W$ is the total vortex energy. The new terms due to spin-transfer are
$\vec{P}_{||} = p_{||} \int_{\rm PC} dV \left( \nabla \Theta \, \sin\Phi + \frac{1}{2} \nabla\Phi \, \sin{2\Theta}\cos\Phi \right),$
which is the component of spin-transfer parallel to the film plane, and
$\vec{P}_{\perp} = p_{\perp} \int_{\rm PC} dV \sin^2\Theta \, \nabla\Phi,$
which is the component of spin-transfer perpendicular to the film plane. The unit vector $(\vec{p}_{||},p_{\perp})$ represents the orientation of the incoming spin-polarization and the subscript $\rm PC$ denotes an integral over the region underneath the point contact. 

In the absence of pinning centers and assuming the sample edges are sufficiently far from the vortex core, the only position-dependent component of the vortex energy arises from the Oersted field generated from the current flow through the point contact. Assuming the current flows through a perfectly cylindrical cross-section in the free layer, we obtain for $W$ in polar coordinates, 
$W = -\mu_0 M_s H_I \int dV f(\vec{r}) \sin\Theta \times \left( \sin\varphi \cos\Phi - \cos{\varphi}\sin\Phi \right)$,
where $f(\vec{r})$ describes the radial variation of the Oersted field amplitude, $H_I$, from the contact center. While it is difficult to evaluate this energy exactly, we find numerically that the functional form $W(\rho) \simeq \kappa I \rho$ gives a reasonably good approximation for realistic $\Theta$ profiles.

The vortex oscillation frequency is found by solving Eq.~\ref{eq:thiele} in the rigid vortex approximation. Assuming the vortex core is sufficiently far from the contact region, we can suppose that the gradient in the polar angle vanishes, $\nabla\Theta = 0$, and the polar angle is given by the uniform tilted angle due to the perpendicular applied field, $\cos{\Theta_0} = H/M_s$, within the contact region. Evaluating the integrals in (\ref{eq:thiele}) within this approximation and using the representation $R_0 \exp(i \psi) = X_0 + i Y_0$, we find the simplified equations of motion,
\begin{align}
\dot{R}_0 + \frac{\alpha D}{G}\dot{\psi} &= \frac{\pi a^2}{2 G R_0^2} \sigma I p_\perp \sin^2\Theta_0,  \\
\dot{\psi} - \frac{\alpha D}{G}\dot{R}_0 &= \frac{1}{G R_0} \frac{\gamma}{M_s} \frac{\partial W}{\partial R_0},
\end{align}
where $G$ and $D$ are constants corresponding to the gyrovector and damping dyadic, respectively. We are interested in the steady state precession for which $\dot{R}_0 = 0$ and $\omega \equiv \dot{\psi}$ is a constant. Using these constraints, we find a vortex precession frequency of
\begin{equation}
\omega = \left(\frac{2}{\pi a^2} \frac{\alpha D \kappa^2}{\sigma p_\perp}\right) \frac{I}{\sin^2{\Theta_0}}.
\label{eq:vortex_freq}
\end{equation}
The dependence on the applied field strength can be introduced by substituting $\theta_0 = \cos^{-1}(H/M_s)$ and approximating the perpendicular spin-polarization component from the $z$ component of the pinned-layer magnetization, $p_\perp \simeq H/M_{2} $, where $M_2$ is the saturation magnetization of the pinned layer.  Note that only the perpendicular component of the spin-polarization leads to vortex precession; the in-plane component only leads  to a displacement of the vortex core. In our system, a non-vanishing contribution from $p_\perp$ arises from the finite tilt angle of the fixed layer magnetization by virtue of the perpendicular applied field.

\begin{figure}
\includegraphics[width=7cm]{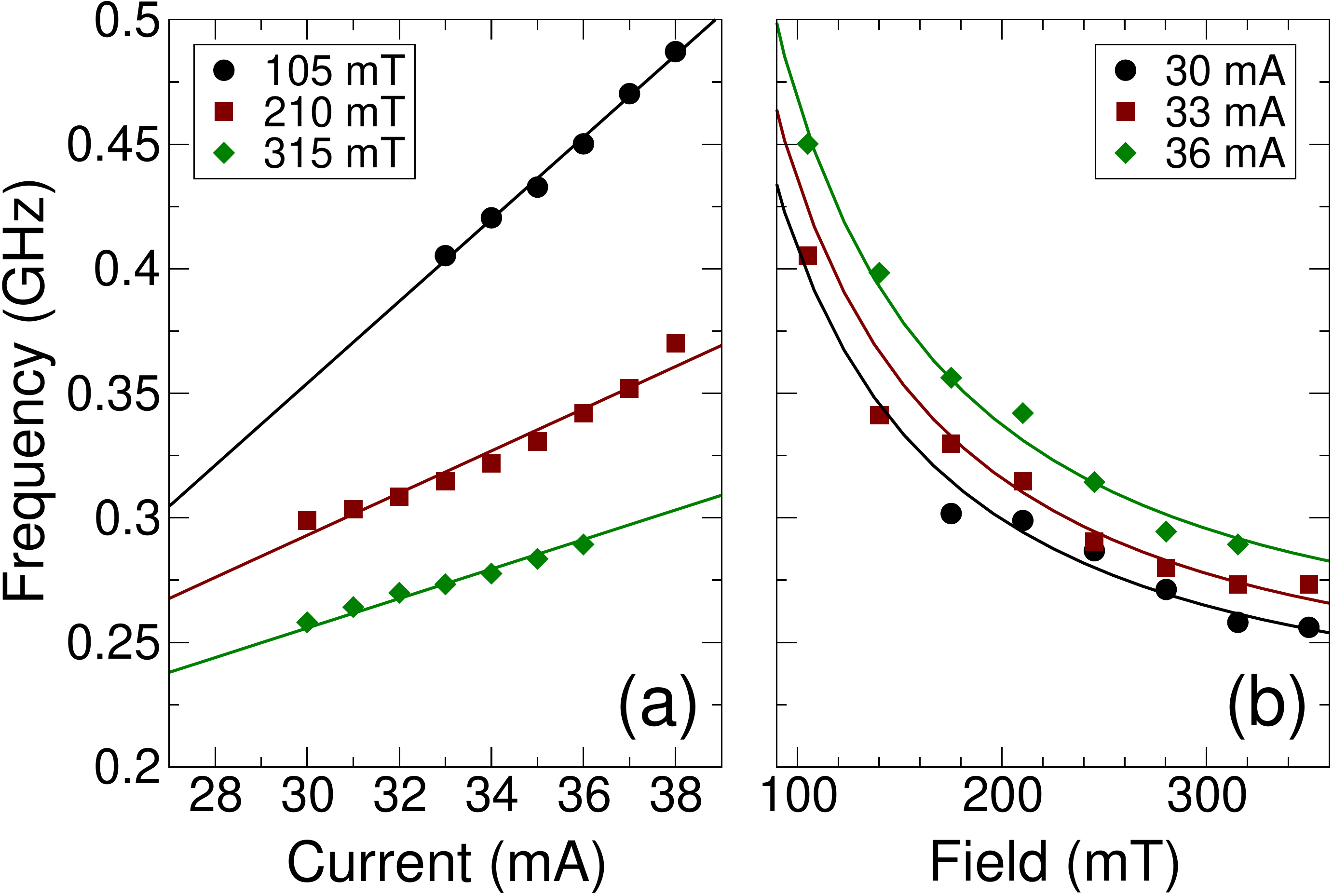}
\caption{\label{fig:expfit}(Color online) Vortex oscillation frequency as a function of (a) applied current and (b) applied field. Dots represent experimental data and solid lines are fits based on Eq.~\ref{eq:vortex_freq}.}
\end{figure}

A comparison between theory and experiment is shown in Fig.~\ref{fig:expfit}.
%
As the current is varied, we observe a quasi-linear variation of the oscillation frequency. This behavior is consistent with theory and is in essence due to the linear dependence of the Oersted field energy on current. To produce the theoretical curves in Fig.~\ref{fig:expfit}a, we first fit the $H =$ 105 mT series using Eq.~\ref{eq:vortex_freq} and then use the theoretical $H$ dependence to compute the slopes for the other applied fields, which are in good quantitative agreement with the experimental data. The theory also accounts for the frequency variation with applied field, as shown in Fig.~\ref{fig:expfit}b. As before, we fit (\ref{eq:vortex_freq}) to one data set (I = 36 mA) and used the predicted current variation for the other sets.

In summary, we have shown with experiment, simulation, and analytical theory that current-driven sub-GHz oscillations in point-contacts are a result of large-amplitude vortex dynamics. The vortex attains a stable circular orbit outside of the contact region, whose frequency is governed by the applied perpendicular field and current.

\begin{acknowledgments}
This work was supported by the European Communities programs IST STREP, under Contract No. IST-016939 TUNAMOS, and ``Structuring the ERA'', under Contract No. MRTN-CT-2006-035327 SPINSWITCH, and by the R{\'e}gion Ile-de-France in the framework of C'nano IdF. Q. M. acknowledges support from a CNRS/STMicroelectronics PhD grant.
\end{acknowledgments}

\bibliography{articles_trunc}

\end{document}